\documentclass[%
reprint,
superscriptaddress,
showpacs,
 amsmath,amssymb,
 aps,prl,
]{revtex4-2}

\usepackage{graphicx}
\usepackage{dcolumn}
\usepackage{bm}
\usepackage{comment}
\usepackage[mathlines]{lineno}

\begin{document}

\preprint{APS/123-QED}

\title{First-time measurement of Timelike Compton Scattering}

\newcommand*{\ANL}{Argonne National Laboratory, Argonne, Illinois 60439}
\newcommand*{\ANLindex}{1}
\affiliation{\ANL}
\newcommand*{\CSUDH}{California State University, Dominguez Hills, Carson, CA 90747}
\newcommand*{\CSUDHindex}{1}
\affiliation{\CSUDH}
\newcommand*{\SACLAY}{IRFU, CEA, Universit\'{e} Paris-Saclay, F-91191 Gif-sur-Yvette, France}
\newcommand*{\SACLAYindex}{2}
\affiliation{\SACLAY}
\newcommand*{\CNU}{Christopher Newport University, Newport News, Virginia 23606}
\newcommand*{\CNUindex}{3}
\affiliation{\CNU}
\newcommand*{\UCONN}{University of Connecticut, Storrs, Connecticut 06269}
\newcommand*{\UCONNindex}{4}
\affiliation{\UCONN}
\newcommand*{\DUKE}{Duke University, Durham, North Carolina 27708-0305}
\newcommand*{\DUKEindex}{5}
\affiliation{\DUKE}
\newcommand*{\DUQUESNE}{Duquesne University, 600 Forbes Avenue, Pittsburgh, PA 15282 }
\newcommand*{\DUQUESNEindex}{6}
\affiliation{\DUQUESNE}
\newcommand*{\FU}{Fairfield University, Fairfield CT 06824}
\newcommand*{\FUindex}{7}
\affiliation{\FU}
\newcommand*{\FERRARAU}{Universita' di Ferrara , 44121 Ferrara, Italy}
\newcommand*{\FERRARAUindex}{8}
\affiliation{\FERRARAU}
\newcommand*{\FIU}{Florida International University, Miami, Florida 33199}
\newcommand*{\FIUindex}{9}
\affiliation{\FIU}
\newcommand*{\FSU}{Florida State University, Tallahassee, Florida 32306}
\newcommand*{\FSUindex}{10}
\affiliation{\FSU}
\newcommand*{\GWUI}{The George Washington University, Washington, DC 20052}
\newcommand*{\GWUIindex}{11}
\affiliation{\GWUI}
\newcommand*{\INFNFE}{INFN, Sezione di Ferrara, 44100 Ferrara, Italy}
\newcommand*{\INFNFEindex}{12}
\affiliation{\INFNFE}
\newcommand*{\INFNFR}{INFN, Laboratori Nazionali di Frascati, 00044 Frascati, Italy}
\newcommand*{\INFNFRindex}{13}
\affiliation{\INFNFR}
\newcommand*{\INFNGE}{INFN, Sezione di Genova, 16146 Genova, Italy}
\newcommand*{\INFNGEindex}{14}
\affiliation{\INFNGE}
\newcommand*{\INFNRO}{INFN, Sezione di Roma Tor Vergata, 00133 Rome, Italy}
\newcommand*{\INFNROindex}{15}
\affiliation{\INFNRO}
\newcommand*{\INFNTUR}{INFN, Sezione di Torino, 10125 Torino, Italy}
\newcommand*{\INFNTURindex}{16}
\affiliation{\INFNTUR}
\newcommand*{\INFNCAT}{INFN, Sezione di Catania, 95123 Catania, Italy}
\newcommand*{\INFNCATindex}{17}
\affiliation{\INFNCAT}
\newcommand*{\INFNPAV}{INFN, Sezione di Pavia, 27100 Pavia, Italy}
\newcommand*{\INFNPAVindex}{18}
\affiliation{\INFNPAV}
\newcommand*{\ORSAY}{Universit\'{e} Paris-Saclay, CNRS/IN2P3, IJCLab, 91405 Orsay, France}
\newcommand*{\ORSAYindex}{19}
\affiliation{\ORSAY}
\newcommand*{\KNU}{Kyungpook National University, Daegu 41566, Republic of Korea}
\newcommand*{\KNUindex}{20}
\affiliation{\KNU}
\newcommand*{\LAMAR}{Lamar University, 4400 MLK Blvd, PO Box 10046, Beaumont, Texas 77710}
\newcommand*{\LAMARindex}{21}
\affiliation{\LAMAR}
\newcommand*{\MIT}{Massachusetts Institute of Technology, Cambridge, Massachusetts  02139-4307}
\newcommand*{\MITindex}{22}
\affiliation{\MIT}
\newcommand*{\MISS}{Mississippi State University, Mississippi State, MS 39762-5167}
\newcommand*{\MISSindex}{23}
\affiliation{\MISS}
\newcommand*{\ITEP}{National Research Centre Kurchatov Institute - ITEP, Moscow, 117259, Russia}
\newcommand*{\ITEPindex}{24}
\affiliation{\ITEP}
\newcommand*{\UNH}{University of New Hampshire, Durham, New Hampshire 03824-3568}
\newcommand*{\UNHindex}{25}
\affiliation{\UNH}
\newcommand*{\NMSU}{New Mexico State University, PO Box 30001, Las Cruces, NM 88003, USA}
\newcommand*{\NMSUindex}{26}
\affiliation{\NMSU}
\newcommand*{\OHIOU}{Ohio University, Athens, Ohio  45701}
\newcommand*{\OHIOUindex}{27}
\affiliation{\OHIOU}
\newcommand*{\ODU}{Old Dominion University, Norfolk, Virginia 23529}
\newcommand*{\ODUindex}{28}
\affiliation{\ODU}
\newcommand*{\JLUGiessen}{II Physikalisches Institut der Universitaet Giessen, 35392 Giessen, Germany}
\newcommand*{\JLUGiessenindex}{29}
\affiliation{\JLUGiessen}
\newcommand*{\URICH}{University of Richmond, Richmond, Virginia 23173}
\newcommand*{\URICHindex}{31}
\affiliation{\URICH}
\newcommand*{\ROMAII}{Universita' di Roma Tor Vergata, 00133 Rome Italy}
\newcommand*{\ROMAIIindex}{30}
\affiliation{\ROMAII}
\newcommand*{\MSU}{Skobeltsyn Institute of Nuclear Physics, Lomonosov Moscow State University, 119234 Moscow, Russia}
\newcommand*{\MSUindex}{31}
\affiliation{\MSU}
\newcommand*{\SCAROLINA}{University of South Carolina, Columbia, South Carolina 29208}
\newcommand*{\SCAROLINAindex}{32}
\affiliation{\SCAROLINA}
\newcommand*{\TEMPLE}{Temple University,  Philadelphia, PA 19122 }
\newcommand*{\TEMPLEindex}{33}
\affiliation{\TEMPLE}
\newcommand*{\JLAB}{Thomas Jefferson National Accelerator Facility, Newport News, Virginia 23606}
\newcommand*{\JLABindex}{34}
\affiliation{\JLAB}
\newcommand*{\UTFSM}{Universidad T\'{e}cnica Federico Santa Mar\'{i}a, Casilla 110-V Valpara\'{i}so, Chile}
\newcommand*{\UTFSMindex}{35}
\affiliation{\UTFSM}
\newcommand*{\INSUBRIA}{Universit\`{a} degli Studi dell'Insubria, 22100 Como, Italy}
\newcommand*{\INSUBRIAindex}{36}
\affiliation{\INSUBRIA}
\newcommand*{\BRESCIA}{Universit\`{a} degli Studi di Brescia, 25123 Brescia, Italy}
\newcommand*{\BRESCIAindex}{37}
\affiliation{\BRESCIA}
\newcommand*{\MESSU}{Universit`{a} degli Studi di Messina, 98166 Messina, Italy}
\newcommand*{\MESSUindex}{38}
\affiliation{\MESSU}
\newcommand*{\GLASGOW}{University of Glasgow, Glasgow G12 8QQ, United Kingdom}
\newcommand*{\GLASGOWindex}{39}
\affiliation{\GLASGOW}
\newcommand*{\YORK}{University of York, York YO10 5DD, United Kingdom}
\newcommand*{\YORKindex}{40}
\affiliation{\YORK}
\newcommand*{\VIRGINIA}{University of Virginia, Charlottesville, Virginia 22901}
\newcommand*{\VIRGINIAindex}{41}
\affiliation{\VIRGINIA}
\newcommand*{\WM}{College of William and Mary, Williamsburg, Virginia 23187-8795}
\newcommand*{\WMindex}{42}
\affiliation{\WM}
\newcommand*{\YEREVAN}{Yerevan Physics Institute, 375036 Yerevan, Armenia}
\newcommand*{\YEREVANindex}{43}
\affiliation{\YEREVAN}

\newcommand*{\NOWMISS}{Mississippi State University, Mississippi State, MS 39762-5167}
\newcommand*{\NOWANL}{Argonne National Laboratory, Argonne, Illinois 60439}
\newcommand*{\NOWBRESCIA}{Universit\`{a} degli Studi di Brescia, 25123 Brescia, Italy}
\newcommand*{\NOWGENOVA}{INFN, Sezione di Genova, 16146 Genova, Italy}

\author{P.~Chatagnon}\email{pierre.chata@orange.fr}
\altaffiliation[Current address: ]{\NOWGENOVA}
\affiliation{\ORSAY}
\author{S.~Niccolai}
\affiliation{\ORSAY}
\author{S.~Stepanyan}
\affiliation{\JLAB}
\author {M.J.~Amaryan} 
\affiliation{\ODU}
\author {G.~Angelini} 
\affiliation{\GWUI}
\author {W.R.~Armstrong} 
\affiliation{\ANL}
\author {H.~Atac} 
\affiliation{\TEMPLE}
\author {C.~Ayerbe Gayoso} 
\altaffiliation[Current address: ]{\NOWMISS}
\affiliation{\WM}
\author {N.A.~Baltzell} 
\affiliation{\JLAB}
\author {L. Barion} 
\affiliation{\INFNFE}
% ...

\collaboration{The CLAS Collaboration}
\noaffiliation

\date{\today}

\begin{abstract}
We present the first measurement of the Timelike Compton Scattering process, $\gamma p\to p^\prime \gamma^* (\gamma^*\to e^+e^-) $, obtained with the CLAS12 detector at Jefferson Lab. The photon beam polarization and the decay lepton angular asymmetries are reported in the range of timelike photon virtualities $2.25<Q^{\prime 2}<9$ GeV$^2$, squared momentum transferred $0.1<-t<0.8$ GeV$^2$, and average total center-of-mass energy squared ${s}=14.5$ GeV$^2$. The photon beam polarization asymmetry, similar to the beam-spin asymmetry in Deeply Virtual Compton Scattering, is sensitive to the imaginary part of the Compton Form Factors and provides a way to test the universality of the Generalized Parton Distributions. The angular asymmetry of the decay leptons accesses the real part of the Compton Form Factors and thus the D-term in the parametrization of the Generalized Parton Distributions.
\end{abstract}

\maketitle

Most of the mass of the observable universe comes from protons and neutrons. The mass of nucleons comes mainly from the interactions between their fundamental constituents, the quarks and the gluons (also referred to as ``partons"), which are described by the Quantum Chromodynamics (QCD) Lagrangian \cite{Altarelli:1981ax}. 
However, QCD-based calculations cannot yet be performed to fully explain the properties of nucleons in terms of their constituents. Therefore, phenomenological functions are used to connect experimental observables with the dynamics of partons in nucleons. Typical examples of such functions are the form factors (FFs) and parton distribution functions (PDFs). Generalized Parton Distributions (GPDs) combine and extend the information contained in FFs and PDFs \cite{Diehl_2016}.  They describe the correlations between the longitudinal momentum and transverse spatial position of the partons inside the nucleon, giving access to the contribution of the orbital momentum of the quarks to the nucleon, and they are sensitive to the correlated $q$-$\bar{q}$ components \cite{muller,ji,PhysRevD.55.7114,PhysRevD.56.5524,Radyushkin:1998bz,PhysRevD.62.071503}. 

\begin{figure}[htb]
\includegraphics[width=\linewidth]{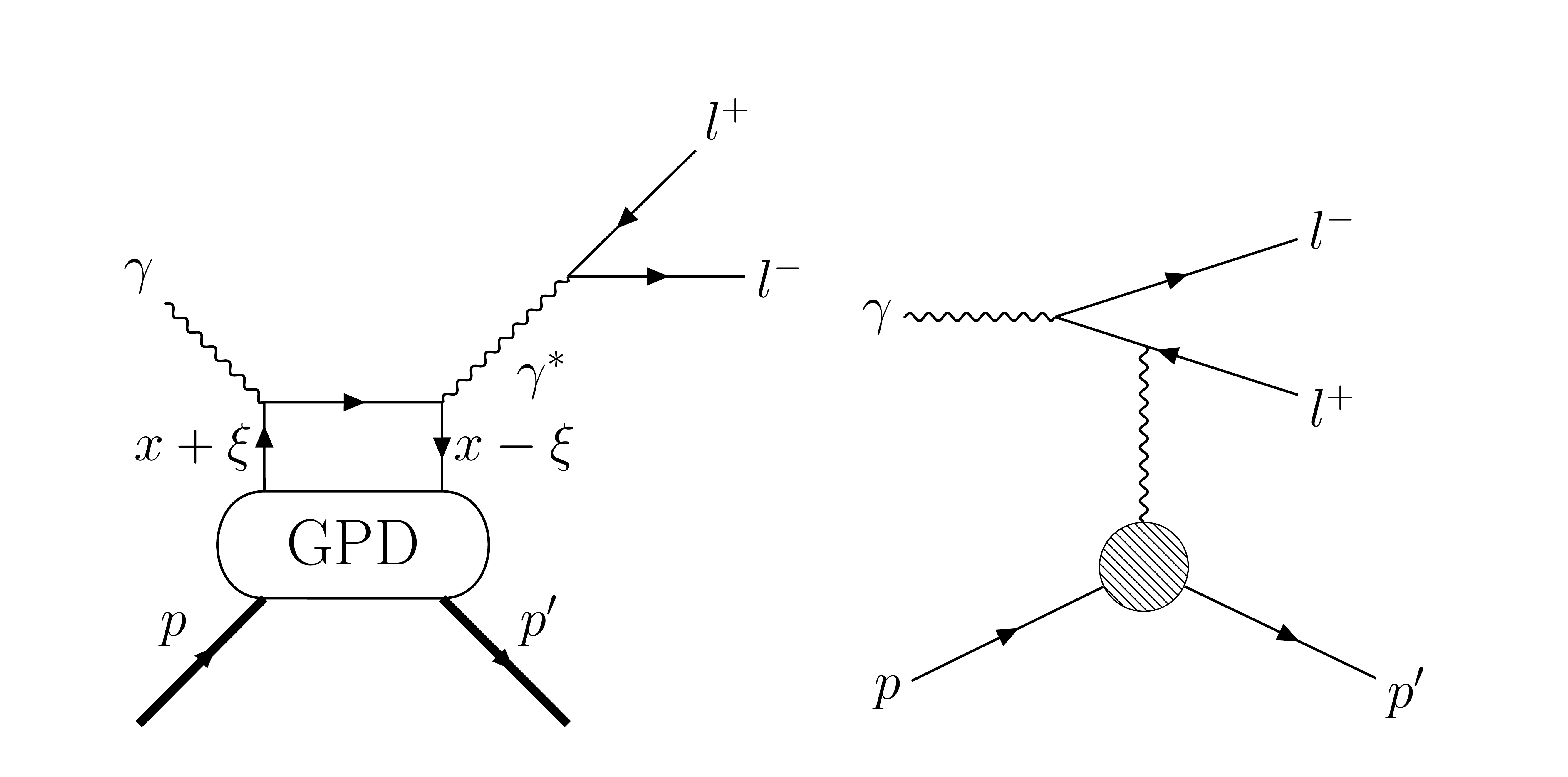}
\caption{\label{fig:TCS} Left: handbag diagram of the TCS process; right:~diagram of the Bethe Heitler (BH) process. $t=(p - p')^2$ is the squared four-momentum transfer between the initial and final protons. $\xi = \frac{\tau}{2-\tau}$ is the momentum imbalance of the struck quark, where $\tau=\frac{Q^{\prime 2 }}{(s-m_p^2)}$, $s$ is the squared center-of-mass energy, and $m_p$ is the proton mass. $x$ is the average momentum fraction of the struck quark.}
\end{figure}

Compton scattering has long been identified as a golden process among deep exclusive reactions to study GPDs experimentally. Deeply Virtual Compton Scattering (DVCS), the exclusive electroproduction of a real photon ($ep\to e^\prime p^\prime\gamma$), has been the preferred tool for accessing GPDs until now \cite{Stepanyan:2001sm,Camacho:2006qlk,Girod:2007aa,PhysRevLett.114.032001,PhysRevD.91.052014,PhysRevLett.115.212003}. Another Compton process, Timelike Compton Scattering (TCS), has been widely discussed theoretically \cite{TCS2002,TCS2015,NadelTuronski:2009zz,refId0} but never measured experimentally. This article reports on the first measurement of TCS on the proton, $\gamma p\to p^\prime \gamma^* (\gamma^*\to e^+e^-$), with quasi-real photon beam. TCS is the time-reversal symmetric process to DVCS: the incoming photon is real and the outgoing photon has large timelike virtuality. In TCS, the virtuality of the outgoing photon, $Q^{\prime 2}\equiv M^2$, where $M$ is the invariant mass of the lepton pair, sets the hard scale. In the regime $\frac{-t}{Q'^2}\ll 1$
, where $t$ is the squared momentum transfer to the target proton, the factorization theorem \cite{Collins_1999} applies (see Fig.~\ref{fig:TCS}, left). The TCS amplitude can then be expressed as a convolution of the hard scattering amplitude with GPDs, appearing in Compton Form Factors (CFFs). At leading order in $\alpha_s$, the CFF for the GPD $H$ 
is defined in Ref.~\cite{TCS2002} using the notations of Ref.~\cite{BELITSKY2001289} as:
\begin{equation}
\label{CFF}
\mathcal{H} (\xi,t) =\int_{-1}^{1}dx ~H(x,\xi,t)  \left( \frac{1}{\xi-x+i\epsilon} - \frac{1}{\xi+x+i\epsilon} \right),
\end{equation}
where $x$, $\xi$, and $t$ are defined in Fig. \ref{fig:TCS}. Similar equations apply to the other GPDs $E$, $\tilde{E}$, and $\tilde{H}$. 
With a beam of circularly polarized photons TCS can access both the real and imaginary parts of the CFFs \cite{TCS2015}.

\begin{figure}[htb]
\includegraphics[width=0.7\linewidth]{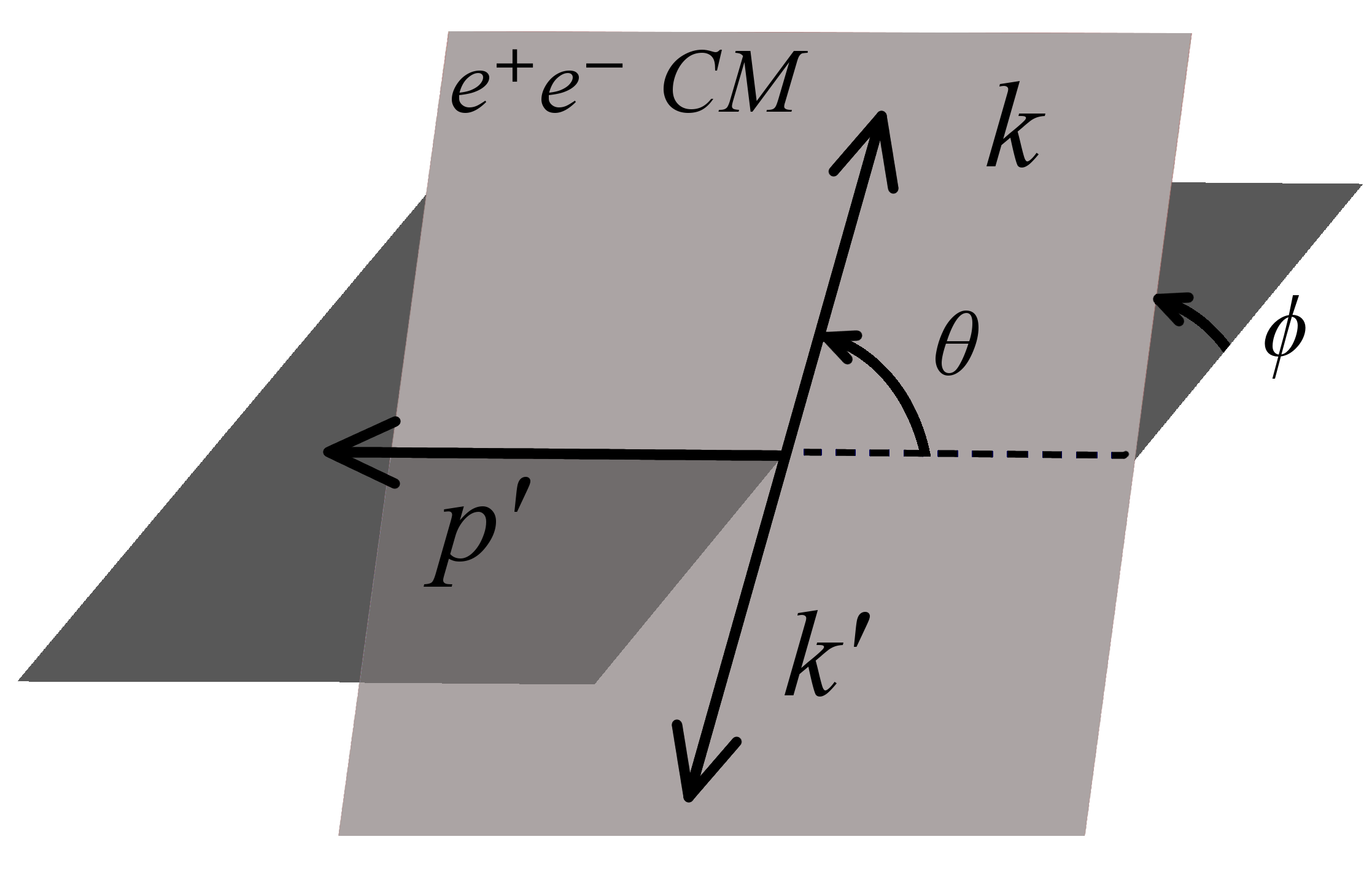}
\caption{\label{fig:Angle} Definition of the relevant angles for the TCS reaction. $\phi$ and $\theta$ are, respectively, the angle between the leptonic plane (defined by the outgoing leptons momenta) and the hadronic plane (defined by the incoming and outgoing proton momenta) , and the angle between the electron and the recoiling proton in the leptons center-of-mass frame.}
\end{figure}

As in the case of DVCS, the Bethe-Heitler process, which can be computed in a quasi-model-independent way, contributes to the same final state (see Fig.~\ref{fig:TCS}, right). 
The cross section for exclusive lepton pair photoproduction on the proton can be expressed as:
\begin{equation}
\sigma(\gamma p \rightarrow p' e^{+}e^{-})=\sigma_{BH}+\sigma_{TCS}+\sigma_{INT},
\end{equation}
where $INT$ stands for the TCS-BH interference term. 
As presented in Ref.~\cite{TCS2002,TCS2015}, the BH contribution dominates over the TCS in the total cross section by two orders of magnitude in the kinematic range accessible at Jefferson Lab (JLab). Therefore, the best practical way to access GPDs with the TCS reaction is to measure observables giving access to the TCS-BH interference. At leading order and leading twist in QCD, $\sigma_{INT}$ can be expressed as a linear combination of GPD-related quantities \cite{TCS2002}: 
\begin{equation}\label{cs_int}
\begin{split}
\frac{d^{4}\sigma_{INT}}{dQ'^{2}dtd\Omega} =  A\frac{1+\cos^2\theta}{ \sin\theta} [ &\cos\phi~{\rm Re}\tilde{M}^{--} \\ 
-\nu \cdot & \sin\phi~{\rm Im}\tilde{M}^{--} ],
\end{split}
\end{equation}
where
\begin{equation}\label{m_mm}
\tilde{M}^{--}=\left[F_{1}\mathcal{H}-\xi(F_{1}+F_{2})\mathcal{\tilde{H}}-\frac{t}{4m_{p}^{2}}F_{2}\mathcal{E}\right], 
\end{equation}
$A$ is a kinematic factor given in Ref.~\cite{TCS2002}, $\phi$ and $\theta$ are defined in Fig.~\ref{fig:Angle}, $\Omega$ is the solid angle defined by $\theta$ and $\phi$, $\nu$ is the circular polarization of the photon beam (equal to $+1$ for right-handed and -1 for left-handed polarization), $m_{p}$ is the proton mass, $F_{1}$ and $F_{2}$ are the electromagnetic form factors, and $\mathcal{H}$, $\mathcal{\tilde{H}}$, and $\mathcal{E}$ are the TCS Compton Form Factors (CFFs) of the $H$, $\tilde{H}$, and $E$ GPDs, respectively, which are given in Eq.~\ref{CFF}. 
The first term, independent of the polarization, is proportional to the real part of the combination of CFFs $\tilde{M}^{--}$. The second, polarization-dependent term is proportional to $\nu$ multiplied by the imaginary part of $\tilde{M}^{--}$.
As the coefficients of $\mathcal{\tilde{H}}$ and $\mathcal{E}$ in Eq.~\ref{m_mm} are suppressed, especially in the kinematics covered at JLab, measuring observables linked to the TCS-BH interference cross section provides access mainly to the real part of the $\mathcal{H}$ CFF. 

In this work, two TCS observables were measured for the first time: the photon polarization asymmetry and the forward-backward asymmetry. The photon polarization asymmetry for circularly polarized beam ($\odot$) and unpolarized target ($U$), defined as:
\begin{equation}
A_{\odot U}=\frac{d\sigma^+ - d\sigma^-}{d\sigma^+ + d\sigma^-},
\end{equation} 
is proportional to the $\sin{\phi}$ moment of the polarized interference cross section and allows access to the imaginary part of $\mathcal{H}$. Here the superscript $+/-$ stands for the right-handed/left-handed circular polarization of the real photon.

The forward-backward asymmetry $A_{FB}$, defined as: 
\begin{equation}\label{afb_def}
A_{FB}(\theta,\phi)=\frac{d\sigma(\theta,\phi)-d\sigma(180^\circ-\theta,180^\circ+\phi)}{d\sigma(\theta,\phi)+d\sigma(180^\circ-\theta,180^\circ+\phi)},
\end{equation}
projects out the $\cos{\phi}$ moment of the unpolarized cross section, proportional to the real part
of the CFF $\mathcal{H}$. This asymmetry has the advantage to remove a potential false asymmetry arising from the integration over the finite angular coverage of the detectors, compared to the cross-section ratio proposed in Ref.~\cite{TCS2002}. 
Both $A_{\odot U}$ and $A_{FB}$ are zero if only BH contributes to the $\gamma p \to p^{\prime} \gamma^*$ cross section. 
Furthermore it was shown in Ref.~\cite{TCS2020} that the QED radiative corrections are negligible for both of these observables.

The experiment was carried out in Hall B at Jefferson Lab, using a 10.6-GeV electron beam, produced by the CEBAF accelerator, impinging on a 5-cm-long liquid-hydrogen target placed at the center of the solenoid magnet of CLAS12 \cite{BURKERT2020163419}. Potential quasi-real photoproduction events ($ep\to p^\prime e^+e^-X$) were selected with one reconstructed electron, one positron, and one proton. The trajectories of charged particles, bent by the torus and solenoid magnetic fields of CLAS12, were measured by the Drift Chambers (DC) and in the Central Vertex Tracker (CVT), providing the charge and momentum of each track. The electrons and positrons were identified combining the information from the High-Threshold Cherenkov counters (HTCC) and the Forward Electromagnetic Calorimeters (ECAL) \cite{REC}. Leptons with momenta below 1 GeV were removed to eliminate poorly reconstructed tracks in the Forward Detector (FD). The background due to positive pions in the positron sample was minimized by means of a neural-network-based multi-variate analysis of transverse and longitudinal profiles of showers in the ECAL. The protons were identified by analyzing the $\beta$ ($\beta = v/c$ where $v$ is the particle's velocity and $c$ the speed of light) of positive tracks measured by the CLAS12 time-of-flight systems (FTOF, CTOF) as a function of their momentum. The momenta of the protons were corrected for energy loss in the detector materials using Monte Carlo simulations. Additional data-driven corrections were included, to account, in the case of the leptons, for radiative losses, and, in the case of protons, for detector-dependent momentum shifts not accounted by the simulation. 

Once the  $p^\prime e^-e^+$ events were selected, exclusivity selection criteria were applied to ensure kinematics in the quasi-real photoproduction regime. The 4-momenta of the scattered electron and initial quasi-real photon were determined via energy-momentum conservation from the measured 4-momenta of the final-state proton and the lepton pair. Then the mass and the transverse momentum fraction $P_t/P$ of the scattered electron were  constrained to be close to zero ($P_t/P<0.05$, $\mid M^2 \mid <0.4$ GeV$^2$). These selection criteria ensure that the virtuality of the incoming photon is low ($Q^2<0.15~{\rm GeV^2}$). In fact, $Q^2$ can be written as:
\begin{equation}
\label{Q2Formula}
Q^2=2E_{b}E_X(1-\cos\theta_X),
\end{equation}
where $E_b$ is the energy of the electron beam, $E_X$ is the energy of the undetected scattered electron and $\theta_X=\arcsin(P_t/P)$ is its scattering angle in the lab frame.

\begin{figure}[htb]
\includegraphics[width=\linewidth]{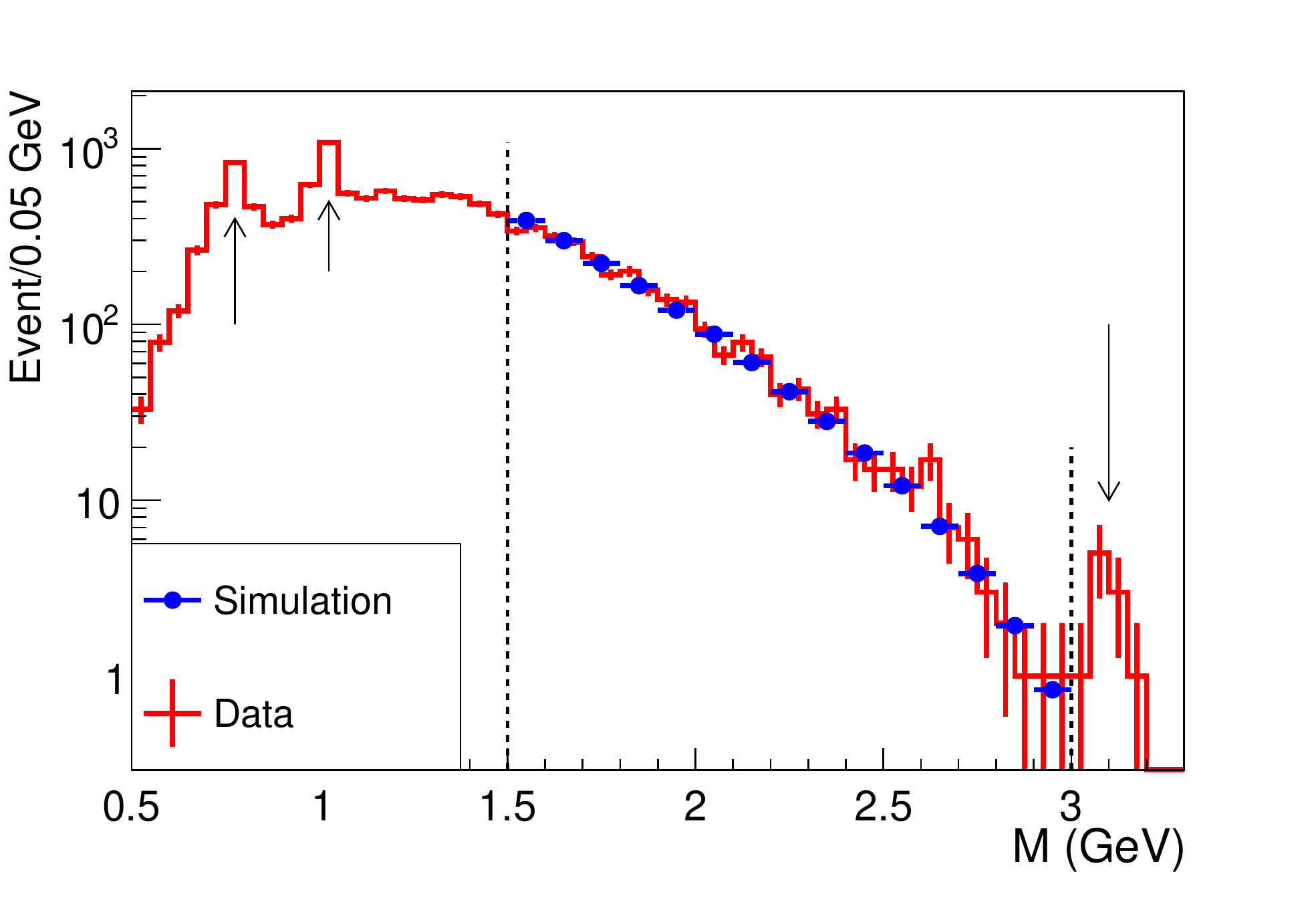}
\caption{\label{fig:Mass} Invariant mass spectrum of the electron-positron pairs. The peaks, indicated by the arrows, correspond to the $\rho_0/\omega$, $\phi$ and $J/\psi$ mesons. The TCS events, represented by the histogram, were selected in the 1.5-3 GeV mass range (within the dotted vertical lines). In this range the data are compared to Monte-Carlo simulation (dots) of Bethe-Heitler events. The simulation is normalized to the total number of events. The data/simulation bin-by-bin ratio agrees at the 15\% level.} 
\end{figure}

The invariant mass spectrum of the outgoing lepton pair after exclusivity selection is shown in Fig.~\ref{fig:Mass}. The vector meson resonances decaying into an electron-positron pair ($\rho_0/\omega$, $\phi$, and $J/\psi$) are clearly visible. 
2921 events with invariant mass between 1.5 GeV and 3 GeV were selected to measure the TCS observables. Indeed in this region the factorization condition $-t/Q'^2\ll 1$ needed for the GPD formalism to apply is fulfilled. In Fig.~\ref{fig:Mass} the experimentally measured invariant mass distribution is compared with BH Monte-Carlo events. The good agreement between the two distributions rules out the possible contamination of the data by high mass meson resonances decaying into $e^+e^-$ pairs ({\it e.g.} $\rho(1450)$ and $\rho(1700)$). 

The photon polarization asymmetry was computed in four bins of $-t$. Each bin has an equal number of events to yield comparable statistical uncertainties. As this analysis is done on quasi-real photoproduction events, where the quasi-real photon is radiated by the initial electron beam, the circular polarization of the photon can be inferred from the initial longitudinal polarization of the electron beam. An electron polarized (with polarization $P_b$) in the direction (opposite) of the beam emits a right-(left-) handed circularly polarized photon, with a transferred polarization  $P_{trans}$ that can be calculated analytically \cite{PhotonPolarization} for each event. Taking advantage of the polarization transfer, the asymmetry $A_{\odot U}$, integrated over $\theta$, is measured as:
\begin{equation}\label{Adotu}
A_{\odot U}(-t,E_{\gamma},M;\phi) = \frac{1}{P_{b}} \frac{N^+ -N^-}{N^+ +N^-},
\end{equation}
where the number of events with reported positive and negative electron helicity in each bin is corrected by the acceptance and efficiency of CLAS12 ($Acc$) for the $\gamma p \rightarrow p^\prime e^- e^+$ reaction, and by the polarization transfer, as:
\begin{equation}
N^\pm=\sum \frac{1}{Acc}  P_{trans}.
\end{equation}

$Acc$ was estimated using the CLAS12 GEANT-4 \cite{AGOSTINELLI2003250} based simulations framework \cite{SIMU}. 
A Monte-Carlo sample of 36 million generated events was used. The acceptance was calculated in a 5-dimensional grid of bins in the variables describing TCS ($-t$, $E_{\gamma}$, $Q^{\prime 2}$, $\theta$, $\phi$). In a given bin, the acceptance is defined as the number of events reconstructed in this bin divided by the number of events generated in this bin.
Low-occupancy bins, yielding an acceptance below 5\% and with a relative uncertainty greater than 50\%, were discarded from the analysis. 

\begin{figure}[htb]
\includegraphics[width=\linewidth]{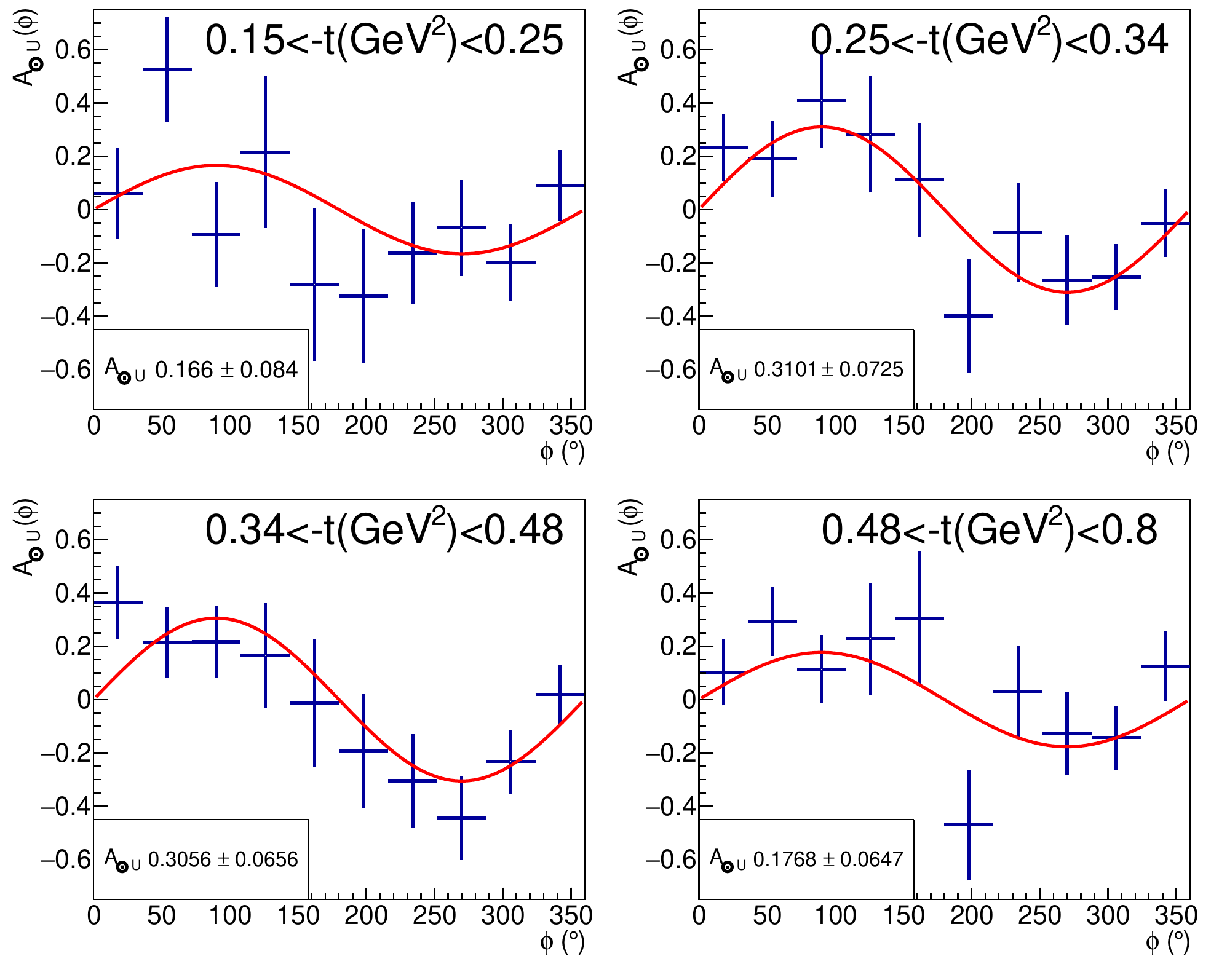}
\caption{\label{fig:Fit} Photon polarization asymmetry as a function of $\phi$ for the four $t$-bins used in this analysis. The sine fit function is superimposed. The amplitude of the fit $A_{\odot U}$ is plotted as a function of $-t$ in Fig.~\ref{fig:BSAvst}. }
\end{figure}

\begin{figure}[htb]\label{fig_bsa_vs_t}
\includegraphics[width=\linewidth]{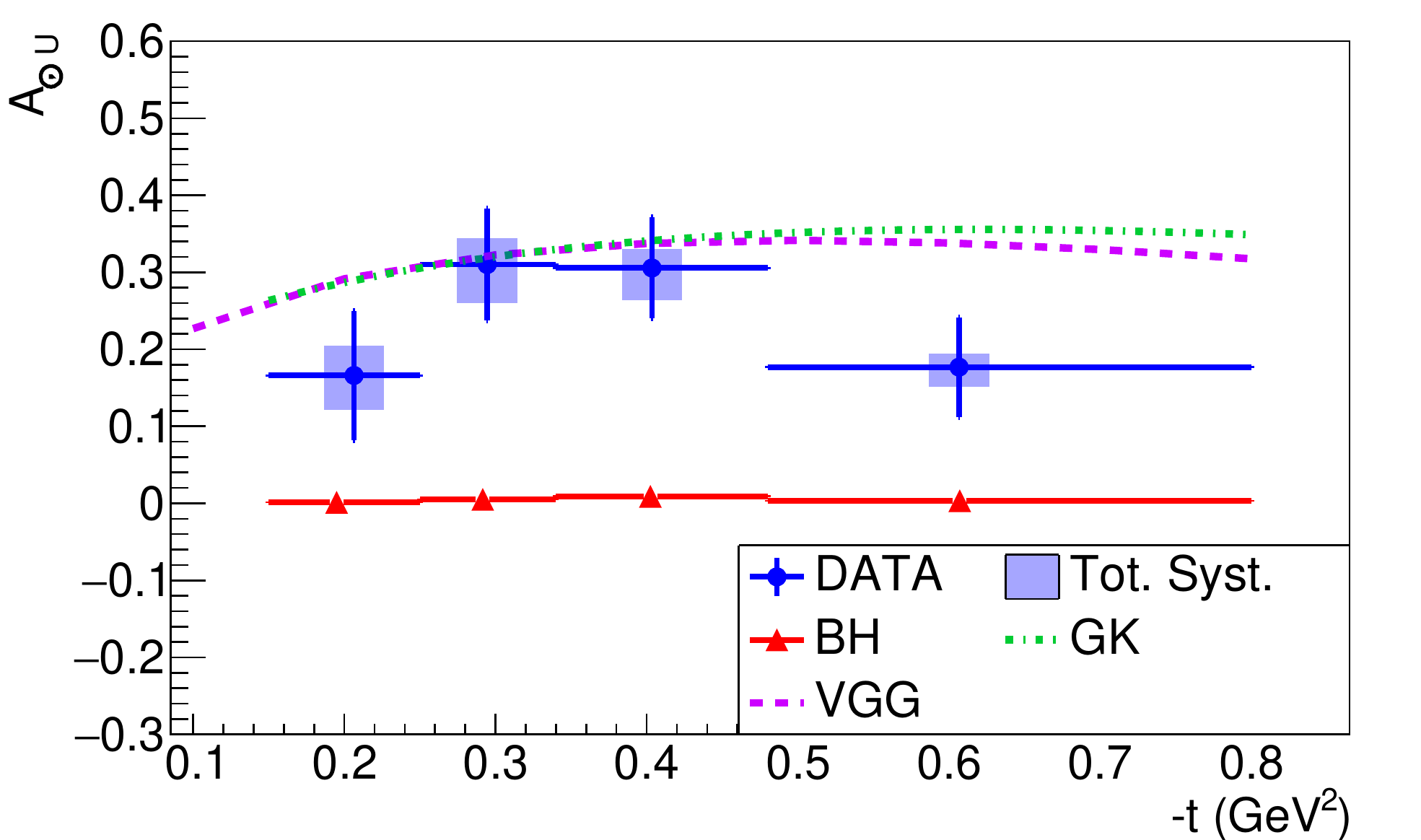}
\caption{\label{fig:BSAvst} Photon polarization asymmetry $A_{\odot U}$ as a function of $-t$ at the averaged kinematic point $E_\gamma = 7.29 \pm 1.55~{\rm GeV}$; $M=1.80\pm 0.26~{\rm GeV}$. The errors on the averaged kinematic point are the standard deviations of the corresponding distributions of events. The data points are represented in blue with statistical vertical error bars. The horizontal bars represent the bin widths. The shaded error bars show the total systematic uncertainty. The red triangles show the asymmetry computed for simulated BH events.  The dashed and dashed-dotted lines are the predictions of, respectively, the VGG \cite{PhysRevLett.80.5064,PhysRevD.60.094017,PhysRevD.72.054013,Guidal:2013rya} and the GK \cite{Goloskokov2005,Goloskokov2008,Goloskokov2009} models, evaluated at the average kinematics.}
\end{figure}

The obtained $\phi$-distributions of the asymmetry of Eq.~\ref{Adotu}  are shown in
Fig.~\ref{fig:Fit}. The distributions are fitted with a sinusoidal function. In Fig.~\ref{fig:BSAvst}, the $-t$ dependence of the amplitude of the sinusoidal modulation is presented.

In-depth systematic checks were performed to validate this measurement. For each identified source of systematic uncertainty, a value of systematic shift was calculated for each bin and added in quadrature after a smoothing procedure. This procedure was necessary to avoid the large fluctuations of the systematic uncertainties from bin-to-bin due to the low statistics of this analysis.
Seven sources of systematic uncertainties were studied: the uncertainties associated with the binning of the acceptance corrections and with the rejection of low-acceptance bins; the uncertainties associated with the Monte Carlo model used to calculate the acceptance and the related efficiency corrections; the systematic shifts induced by the identification procedure of protons and positrons; the impact of the variation of the exclusivity selection criteria. The total systematic uncertainties, given by the quadratic sum of all contributions, are always smaller than the statistical uncertainties, typically by more than 50\%. The major contribution to the systematic uncertainties comes from the exclusivity selection. 

In Figs.~\ref{fig:Fit} and~\ref{fig:BSAvst}, a clear photon beam polarization asymmetry is observed. This agrees with the expected contribution of the BH-TCS interference term to the cross section as the expected asymmetry for the BH contribution only, which was estimated using BH-only Monte-Carlo simulation, is zero.
The photon polarization asymmetries were compared to predictions of the VGG model (based on a double-distribution parametrization with Regge-like $t$-dependence) \cite{PhysRevLett.80.5064,PhysRevD.60.094017,PhysRevD.72.054013,Guidal:2013rya} and of the GK model (based on a double-distribution paramerization with $t$-dependence expressed in the forward limit) \cite{Goloskokov2005,Goloskokov2008,Goloskokov2009} computed within the \textit{PARTONS} framework \cite{PARTONS}. Both of these calculations were performed at leading order in $\alpha_s$, which is a reasonable approximation in our kinematics, while QCD corrections have been shown to be quite important at lower values of $\xi$ \cite{PhysRevD.83.034009,PhysRevD.86.031502,PhysRevD.87.054029}. The measured values are in approximate agreement with the predictions of GPD-based models, while BH-only calculations show no asymmetry. This observation validates the application of the GPD formalism to describe TCS data and hints at the universality of GPDs, as the VGG and GK models also describe well the 6-GeV DVCS data from JLab \cite{Dupre:2016mai}.

Using the same data set, the FB asymmetry, defined in Eq.~\ref{afb_def}, was measured for four bins in $-t$, integrating over all other kinematic variables due to the limited statistics of the event sample. Moreover, the angular coverage of CLAS12 allowed us to measure $A_{FB}$ only in a limited angular range. Thus, the forward and backward angles ($\phi_F$, $\theta_F$, $\phi_B$, $\theta_B$, with $\phi_B=180^\circ + \phi_F$ and $\theta_B=180^\circ-\theta_F$) were extracted in a forward region defined by $-40^\circ<\phi_F<40^\circ$, $50^\circ<\theta_F<80^\circ$, and in a corresponding backward region ($B$) defined by $140^\circ<\phi_B<220^\circ$, $100^\circ<\theta_B<130^\circ$. 
The value of $A_{FB}$ was computed, for each $-t$ bin, as:
\begin{equation}
A_{FB}=\frac{N_F-N_B}{N_F+N_B},
\end{equation}
where $N_{F/B}$ are the number of events in the forward/backward angular bins, corrected by the acceptance and the bin volume. The bin volume correction accounts for the difference in coverage between the forward and the backward directions, that could induce false asymmetries. This correction assumes that the cross section of the TCS reaction is constant within the volume of the forward (resp. backward) bin and that it can be estimated only by measuring it in the volume covered by the acceptance of CLAS12. These approximations were accounted for in the systematic uncertainties by computing $A_{FB}$ with BH-weighted simulated events. The difference between the expected value (null asymmetry, as the BH cross section is symmetric in $\phi$ around $180^{\circ}$) and the obtained value was then assigned as a systematic uncertainty.

\begin{figure}[htb]\label{afb_vs_t}
\includegraphics[width=\linewidth]{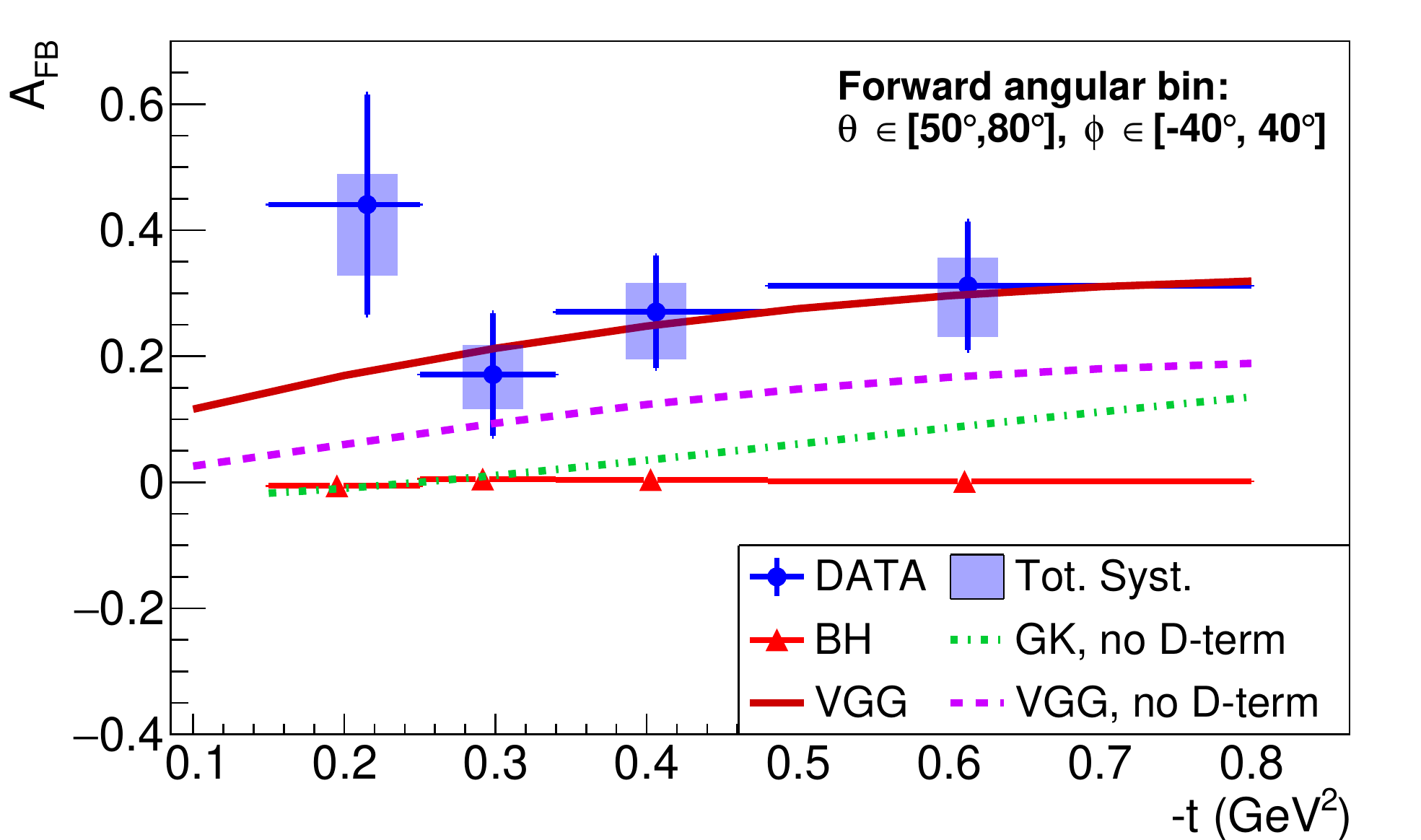}
\caption{\label{fig:AFBt} FB asymmetry as a function of $-t$ at the average kinematics $E_\gamma = 7.23 \pm 1.61~{\rm GeV}$; $M=1.81\pm 0.26~{\rm GeV}$. The solid line shows the model predictions of the VGG model with D-term (from Ref.~\cite{PASQUINI2014133}) evaluated at the average kinematic point. The other curves are defined in the caption of Fig.~\ref{fig_bsa_vs_t}.}
\end{figure}

\begin{figure}[htb]
\includegraphics[width=\linewidth]{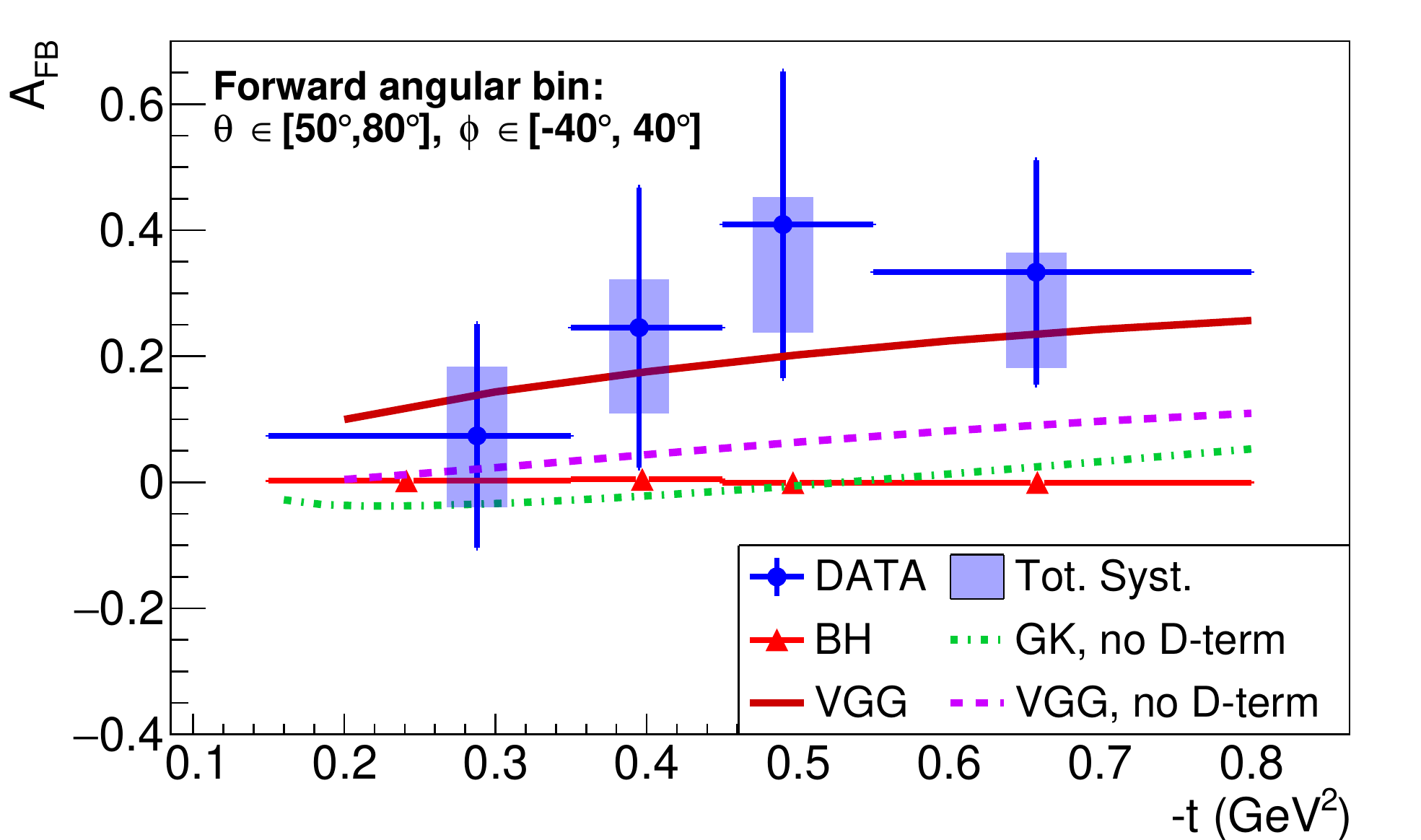}
\caption{\label{fig:AFBt2} FB asymmetry as a function of $-t$ at the average kinematics $E_\gamma = 8.13 \pm 1.23~{\rm GeV}$; $M=2.25\pm 0.20~{\rm GeV}$. The curves are defined in the captions of Figs.~\ref{fig_bsa_vs_t} and \ref{afb_vs_t}.}
\end{figure}

Figure \ref{fig:AFBt} shows $A_{FB}$ for $1.5<M<3$ GeV. In order to explore the dependence on the hard scale ($Q^{\prime 2}\equiv M^2$) of the FB asymmetry, it was extracted separately for the lepton invariant mass region $2$ GeV to $3$ GeV. The results for the high-mass region are shown in Fig.~\ref{fig:AFBt2}. The asymmetries in both mass regions are not comparable with the zero asymmetry predicted if only the BH process was contributing to the total cross section. This confirms that the TCS diagram contributes to the $\gamma p \rightarrow p^\prime e^+ e^-$ cross section.
The experimental results were compared with model predictions. The asymmetries seem to be better described by the VGG model when the D-term (taken from Ref.~\cite{PASQUINI2014133}) is included, although the error bars are still too large to completely rule out the case without the D-term. The D term, a poorly known element of GPD parametrizations that appears as a subtraction term in dispersion relations of DVCS amplitudes, has recently gained relevance for its links to the mechanical properties of the nucleon \cite{POLYAKOV200357,Burkert:2018bqq,Kumericki2019,Dutrieux2021}. The GK model predictions largely underestimate the asymmetry in both mass regions. This could be in part explained by the absence of the D-term in this prediction, although GK differs also from VGG without the D-term. The comparison was also done in the high-mass region in Fig.~\ref{fig:AFBt2}. In this region, where factorization-breaking terms are more strongly suppressed, the previous conclusion stands, supporting the interpretation in terms of GPDs and the importance of the D-term in their parametrization. 

In summary, we reported in this letter the first ever measurement of Timelike Compton Scattering on the proton. Both the photon circular polarization and forward/backward asymmetries were measured. The asymmetries are clearly non-zero, providing strong evidence for the contribution of the quark-level mechanisms parametrized by GPDs to the cross section of this reaction. The comparison of the measured polarization asymmetry with model predictions points toward the interpretation of GPDs as universal functions. Furthermore, the reported results on the FB asymmetry open a new promising path toward the extraction of the real part of $\mathcal{H}$, and ultimately to a better understanding of the internal pressure of the proton via the extraction of the D-term. Future measurements of TCS at JLab will provide a wealth of data to be included in the ongoing fitting efforts to extract CFFs \cite{Kumericki:2016ehc,Dupre:2017hfs,Moutarde_2018,Moutarde2019}. In particular, TCS measurements should have a strong impact in constraining the real part of CFFs \cite{Grocholski2020} and in the determination of the D-term that relates to the gravitational form factor of the nucleon. A comparison of these results with possible measurements of TCS at the EIC \cite{khalek2021science} and in ultra-peripheral collisions at the LHC \cite{PhysRevD.79.014010} could provide a better understanding of the behaviour of the CFFs of TCS at low $x$ \cite{PhysRevD.86.031502,PhysRevD.87.054029}.

We thank Profs. M. Vanderhaeghen, B. Pire, and P. Sznajder for the fruitful exchanges and discussions on the phenomenological aspects of this work and for providing us with the model predictions.
We acknowledge the great efforts of the staff of the Accelerator and the Physics Divisions at Jefferson Lab in making this experiment possible. This work is supported in part by the U.S. Department of Energy, the National Science Foundation (NSF), the Italian Istituto Nazionale di Fisica Nucleare (INFN), the French Centre National de la Recherche Scientifique (CNRS), the French Commissariat pour l'Energie Atomique, the UK Science and  Technology  Facilities  Council,  the  National Research Foundation (NRF) of Korea, the Helmholtz-Forschungsakademie Hessen für FAIR (HFHF) and the Ministry of Science and Higher Education of the Russian Federation.  The  Southeastern  Universities  Research Association  (SURA)  operates  the  Thomas  Jefferson National Accelerator Facility for the U.S. Department of Energy under Contract No. DE-AC05-06OR23177.

\bibliography{TCSarticle}
\bibliographystyle{apsrev4-1}
\end{document}